\documentclass[trackchanges, twocolumn]{aastex701}

\begin{document}

\title{Periodic 6.7 GHz $\mathrm{CH_3OH}$ maser emission in G353.273+0.641: First candidate for a pulsating high-mass protostar}

\author[gname=Sohta, sname=Harajiri]{Sohta Harajiri}
\affiliation{Graduate School of Sciences and Technology for Innovation, Yamaguchi University, Yoshida 1677-1, Yamaguchi 753-8512, Japan}
\email{e022vbv@yamaguchi-u.ac.jp}

\correspondingauthor{Kazuhito Motogi}
\author[orcid=0000-0002-3789-770X, gname=Kazuhito, sname=Motogi]{Kazuhito Motogi}
\affiliation{Graduate School of Sciences and Technology for Innovation, Yamaguchi University, Yoshida 1677-1, Yamaguchi 753-8512, Japan}
\affiliation{The Research Institute for Time Studies, Yamaguchi University, Yoshida 1677-1, Yamaguchi 753-8511, Japan}
\email{kmotogi@yamaguchi-u.ac.jp}

\author[gname=Ryota, sname=Naksmura]{Ryota Nakamura}
\affiliation{Graduate School of Sciences and Technology for Innovation, Yamaguchi University, Yoshida 1677-1, Yamaguchi 753-8512, Japan}
\email{f020vbw@yamaguchi-u.ac.jp}

\author[orcid=0000-0001-5615-5464, gname=Yoshinori, sname=Yonekura]{Yoshinori Yonekura}
\affiliation{Center for Astronomy, Ibaraki University, 2-1-1 Bunkyo, Mito, Ibaraki 310-8512, Japan}
\email{yoshinori.yonekura.sci@vc.ibaraki.ac.jp}

\author[orcid=0000-0002-1183-7899, gname=Yoshinori, sname=Tanabe]{Yoshihiro Tanabe}
\affiliation{The Research Institute for Time Studies, Yamaguchi University, Yoshida 1677-1, Yamaguchi 753-8511, Japan}
\email{oriona@yamaguchi-u.ac.jp}

\author[orcid=0009-0008-1070-4411, gname=Kenta, sname=Fujisawa]{kenta Fujisawa}
\affiliation{The Research Institute for Time Studies, Yamaguchi University, Yoshida 1677-1, Yamaguchi 753-8511, Japan}
\email{kenta@yamaguchi-u.ac.jp}

\begin{abstract}
We report on the periodic flux variations in the 6.7 GHz $\mathrm{CH_3OH}$ maser associated with the high-mass protostar G353.273+0.641, based on 13 yr of monitoring mainly by the Hitachi 32 m telescope. 
We identified a periodicity of 309 days based on a nearly complete light curve, with 833 epochs every few days.
A strong correlation is found between the maser and the mid-infrared fluxes at 3.4 and 4.6 $\mu$m observed by NEOWISE during these periods, suggesting that the maser emission responds to variations in the protostellar luminosity.
The average profile of the maser light curve is asymmetric and shows a steep drop in intensity just before the brightening, resembling that of some pulsating variable stars.
Assuming a protostellar pulsation as the origin of maser periodicity, the observed period implies a cool and highly bloated, red supergiant-like structure.
Such a bloated structure is consistent with a theoretical model of protostellar evolution under high accretion rates.
The inferred protostellar parameters are broadly consistent with the theoretical model of pulsational instability during the early phase of high-mass star formation.
However, a periodic accretion scenario caused by an unresolved compact protobinary cannot be completely ruled out.
Several irregular peaks that deviate from the periodicity may result from episodic accretion phenomena or jet-launching events independent of the protostellar pulsation.
Extremely high-resolution imaging with next-generation interferometers such as the ngVLA will provide a conclusive test for both the protostellar pulsation and the binary accretion scenarios.

\end{abstract}



\section{Introduction}
In recent years, high-resolution observations by the Atacama Large Millimeter/submillimeter Array (ALMA) have been revealing the stability of accretion disks around a high-mass protostar \citep{Beltr_2016A&ARv..24....6B,Girart_2018ApJ...856L..27G,Motogi_2019ApJ...877L..25M,Cesaroni_2025A&A...693A..76C}. 
ALMA has also found several high-mass protobinary systems, and some of which are likely formed via disk fragmentation \citep{Zhang_2019NatAs...3..517Z}. 
While, observational information within 100 au of the central protostar is still limited \citep{Hirota_2017NatAs...1E.146H,Ginsburg_2019ApJ...872...54G, Ginsburg_2023ApJ...942...66G}.
Class II 6.7 GHz $\mathrm{CH_3OH}$ masers, which are commonly associated with the disks and inner envelopes around high-mass protostars, offer a valuable alternative method for studying the innermost accretion activities 
\citep{Sanna_2010A&A...517A..78S, Goddi_2011A&A...535L...8G, Moscadelli_2011A&A...536A..38M, Sugiyama_2014A&A...562A..82S, Motogi_2017ApJ...849...23M, Burns_2023NatAs...7..557B, Nakamura_2023MNRAS.526.1000N}.
This maser corresponds to the $5_1$--$6_0A^+$ transition \citep{Menten_1991ApJ...380L..75M} and is excited by mid-infrared (MIR) radiation \citep{Cragg_2005MNRAS.360..533C}. 
Because of its radiative pumping nature, the maser flux can vary in response to changes in infrared emission from the host protostar. 
Therefore, the maser variability is a potential tracer of the variable protostellar luminosity, and consequently, its accretion rate. 

Explosive increases in 6.7 GHz $\mathrm{CH_3OH}$ maser luminosity have been confirmed in objects such as
S255IR-NIRS3 \citep{Fujisawa_2015ATel.8286....1F}, NGC6334I-MM1 \citep{Hunter_2017ApJ...837L..29H,Hunter_2018ApJ...854..170H}, G358.9-0.03 \citep{Sugiyama_2019ATel12446....1S}, and G24.33+0.14 \citep{Hirota_2022PASJ...74.1234H}.
These observations suggest that the increase in $\mathrm{CH_3OH}$ maser luminosity is due to the rapid increase of the accretion rate as fragmented gas clumps in the accretion disk fall onto the central star (so-called accretion burst).

On the other hand, over 30 periodically variable 6.7 GHz $\mathrm{CH_3OH}$ maser sources have been identified (\citealt[][and references therein]{Tanabe_2024PASJ...76..426T}; 
\citealt{Szymczak_2024A&A...682A..17S}; 
\citealt{Wolak_2025A&A...701A..28W}).
For example, the archetypal object G9.62+0.20 showed periodic and intermittent variations in the maser flux with a period of approximately 243 days \citep{Goedhart_2003MNRAS.339L..33G}.
This behavior has been explained by the colliding-wind binary (CWB) model, in which the enhanced free-free seed photons produced by colliding stellar winds near the periastron drive the periodic maser variability \citep{vanderWalt_2011AJ....141..152V}. 
We note that some other models, such as a protobinary interaction \citep{Parfenov_Sobolev_2014MNRAS.444..620P} or protostellar pulsation \citep{Sanna_2015ApJ...804L...2S}, are still under consideration, although the CWB model appears most plausible in G9.62+0.20 \citep{vanderWalt_2016A&A...588A..47V}. 
In addition to the models that explain the origin of periodicity,\citet{Rajabi_2023MNRAS.526..443R} and \citet{Rashidi_2025MNRAS.542L..12R} proposed that Dicke's superradiance may shape the short, burst-like brightening during each flare.

Although, multiple physical origins are discussed for several periodic $\mathrm{CH_3OH}$ maser sources, in general, these episodic or periodic variations in maser flux are mostly believed to reflect changes in radiation from the protostar and surrounding gas/dust. 

The target source in this study, G353.273+0.641 (hereafter G353), is a relatively nearby high-mass protostar in the southern sky \citep[1.7 kpc;][]{Neckel_1978A&A....69...51N,Motogi_2016PASJ...68...69M}, located at R.A. (J2000.0) = $17^{\mathrm{h}}26^{\mathrm{m}}01^{\mathrm{s}}.59$ and Decl. (J2000.0) = $-34^\circ15'14''.9$.
The 6.7 GHz $\mathrm{CH_3OH}$ masers in G353 have been detected inside the nearly face-on disk within 100 au from the protostar \citep{Caswell_Phillips_2008MNRAS.386.1521C, Motogi_2017ApJ...849...23M}.
\citet{Song_2025ApJ...980..132S} reported a periodic maser flux variation of $\sim$ 330 days for this maser based on 710 days of monitoring observations with the Tianma 65 m radio telescope during 2022--2023.
In this paper, we report on further detailed analysis of this flux variation, based on 13 yr of monitoring observations conducted with the Hitachi 32m and Yamaguchi 32m radio telescopes. 

\section{Observations} \label{sec:style}
Monitoring observations of the 6.7 GHz $\mathrm{CH_3OH}$ maser in G353 were carried out using the Hitachi 32m radio telescope \citep[Ibaraki University; ][]{Yonekura_2016PASJ...68...74Y} and the Yamaguchi 32m radio telescope \citep[Yamaguchi University; ][]{Fujisawa_2002aprm.conf....3F}, in order to investigate the maser flux variations.
At the Hitachi radio telescope, monitoring observations began in January 2013.
The observational cadence is once every $\sim 10$ days from the start of the monitoring observations to 2015 August, and once every $\sim 5$ days from 2015 September to the present (2025 May 30).
At the Yamaguchi radio telescope, daily observations began in October 2024.
Table \ref{tab:obs_sam} shows observational summaries.

Observations at both radio telescopes were conducted using a position-switching method.
The OFF position was set to $\Delta R.A.=+60'$ from the G353.
The integration time per observation is 5 minutes for both the ON and OFF positions.
A single left circular polarization (LCP) signal was recorded at 64 Mbps (16 mega-samples per second with 4 bit sampling) by using a K5/VSSP32 sampler \citep{K5VSSP32_2008}. 
The recorded bandwidth is 8 MHz (RF: 6664 -- 6672 MHz) and they are divided into 8192 channels.
After moving averaging every three channels, the 1-sigma root-mean-squares (rms) noise level is $\sim$ 0.3 Jy for Hitachi radio telescope and 0.5 Jy for the Yamaguchi radio telescope, with a velocity resolution of $\sim$ 0.13 km s$^{-1}$.
The half-power beam width of the Hitachi radio telescope is $\sim4'.6$ with better pointing accuracy than $\sim 30''$ \citep{Yonekura_2016PASJ...68...74Y}.
The half-power beam width of the Yamaguchi radio telescope is $\sim5'.0$, and its pointing error is also within $\sim 30''$.
Although the impact of these errors on the amplitude is relatively small (less than $\sim 3\%$), five-point cross-scan observations were conducted for all measurements with the Yamaguchi telescope to improve data accuracy.
Both radio telescopes measured the antenna temperature using the chopper wheel method.
We used aperture efficiencies ($\eta_\mathrm{A}$) of 0.7 for the Hitachi telescope and 0.6 for the Yamaguchi telescope.
A comparison of flux densities from simultaneous observations at both radio telescopes showed good agreement within a 3-$\sigma$ range, indicating consistent amplitude calibration between the two telescopes.

\begin{deluxetable*}{llcc}
\tablecaption{Observational Summaries \label{tab:obs_sam}}
\tablewidth{0pt}
\tablehead{
\colhead{} & \colhead{} & \colhead{Hitachi 32m} & \colhead{Yamaguchi 32m}
}
\startdata
Frequency [MHz] & & 6664--6672 & 6664--6672 \\
Bandwidth [MHz] & & 8 & 8 \\
Spectral channels & & 8192 & 8192 \\
Channel spacing [km\,s$^{-1}$] & & 0.044 & 0.044 \\
Velocity resolution [km\,s$^{-1}$] & & 0.13 & 0.13 \\
On source time [sec] & & 300 & 300 \\
$\eta_\mathrm{A}$ && 0.7& 0.6 \\
$T_{\mathrm{sys}}$ [K] & & 25--55 & 45--65 \\
1$\sigma$ rms [Jy] & & $\sim$ 0.3 & $\sim$ 0.5 \\
\enddata
\end{deluxetable*}

\section{NEOWISE Archival Data} \label{sec:style}
In this study, we utilized archival data obtained by NASA's NEOWISE mission to investigate the relationship between long-term infrared variations and maser flux.
NEOWISE is a project that reuses NASA's WISE (Wide-field Infrared Survey Explorer) and has been continuously conducting infrared observations of Near-Earth Objects (NEOs) since 2013 \citep{Mainzer_2014ApJ...792...30M}. 
NEOWISE offers two mid-infrared bands at 3.4 $\mu$m (W1 band) and 4.6 $\mu$m (W2 band). 

We acquired time-series data for G353 from the NEOWISE-R Single Exposure (L1b) Source Table for variability analysis\footnote{\url{ http://wise2.ipac.caltech.edu/docs/release/neowise/}}.
We set a search radius of 10 arcsec when associating the NEOWISE source with the maser position.
The coordinates of the NEOWISE infrared source coincide with the millimeter continuum peak of G353 within 1 arcsec.
For the photometric data, we applied criteria to ensure reliability: $1\sigma \leq 0.1 \ \mathrm{mag}$, Signal-to-Noise Ratio (SNR) $\geq 5$, and a high-quality flag of `AA' (ph\_qual).
Additionally, we averaged all the scans obtained on the same day and adopted the standard error of the mean as the uncertainty. 
The average and range of the standard errors were 0.30 (0.08 -- 0.75) mag for the W1 band and 0.43 (0.09 -- 1.60) mag for the W2 band. 

\section{Results} 
\subsection{6.7 GHz $\mathrm{CH_3OH}$ maser light curve}
\begin{figure}[h]
    \centering
    \includegraphics[width=1\linewidth]{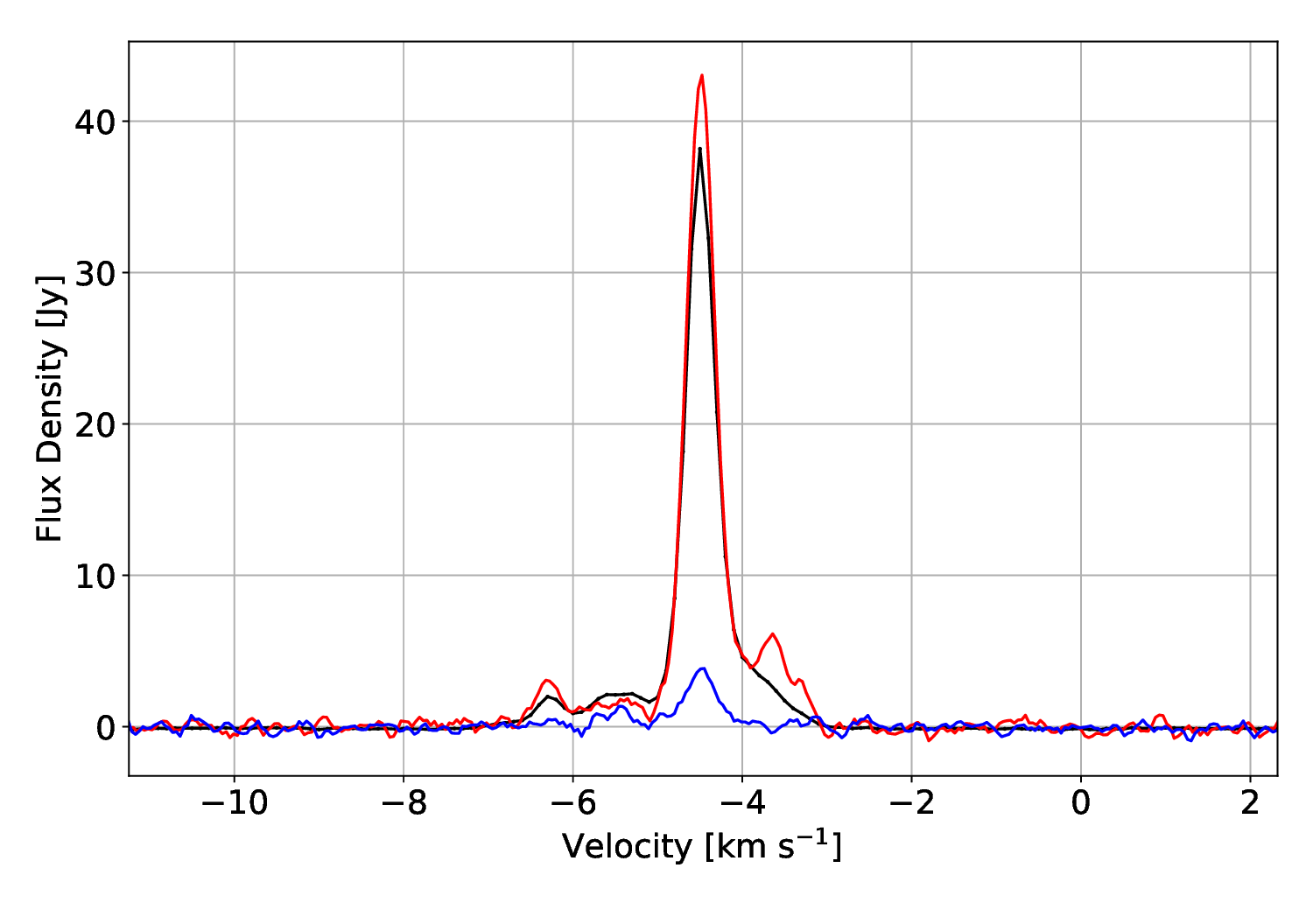}
    \caption{Averaged 6.7 GHz $\mathrm{CH_3OH}$ maser spectrum constructed only using channels $\geq\ 3\sigma$ (black). The spectra from the epochs of maximum (red) and minimum (blue) peak flux are also overlaid to show the range of variability.
}

    \label{fig:average_spec}
\end{figure}

Figure \ref{fig:average_spec} presents the averaged maser spectrum, with spectra at the epochs of maximum and minimum peak flux overlaid to illustrate the range of variability. 
Here we computed the average flux for each channel only using the data $\geq\ 3\sigma$ from the whole epoch \citep[e.g.,][]{Felli_2007A&A...476..373F}. 

\begin{figure*}[h]
    \centering
    \includegraphics[width=0.9\linewidth]{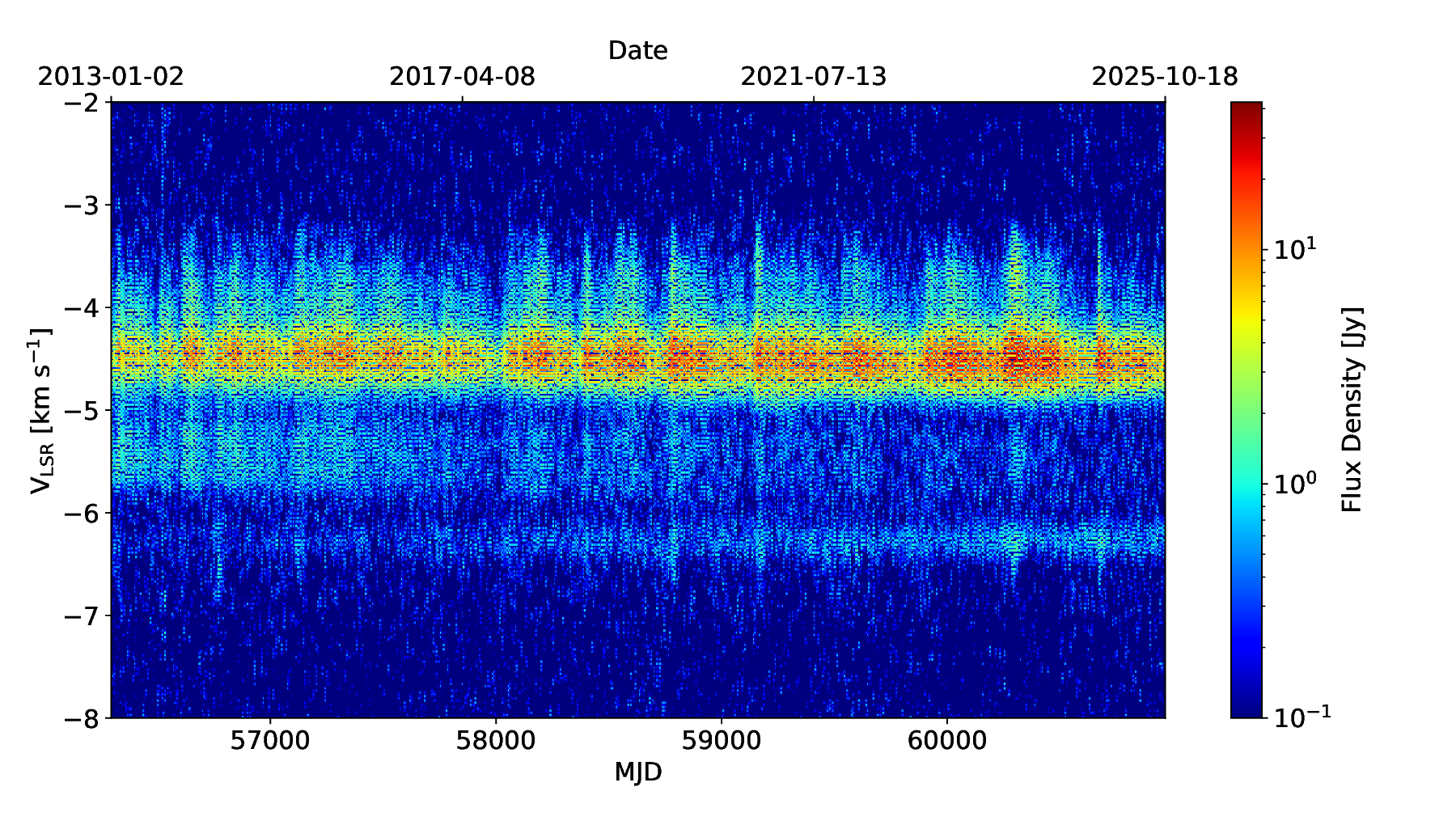}
    \caption{Intensity of the 6.7 GHz $\mathrm{CH_3OH}$ maser as a function of LSR velocity and modified Julian date of the observations. The plot covers the whole monitoring period. No emission was detected outside the displayed velocity range.}
    \label{fig:dynamic}
\end{figure*}

\citet{Motogi_2017ApJ...849...23M} conducted observations with the Australia Telescope Compact Array (ATCA), resolving 43 maser spots and revealing the systematic velocity gradients ($V_\mathrm{LSR}=-6.5 $--$ -3.5\ \mathrm{km\ s^{-1}}$) along the spatial distribution. 
The spectrum in Figure \ref{fig:average_spec} also features multiple overlapping components within a similar velocity range. 
We identified components at $V_\mathrm{LSR}=-3.7$ and $-6.4\ \mathrm{km\ s^{-1}}$. 
These two components are the only sub-peaks consistently detected over the 13-year monitoring, while all other sub-peaks occasionally appeared. 
We will also discuss the flux variations of these components along with the main peak later.
The component at $V_\mathrm{LSR} = -5.5\ \mathrm{km\ s^{-1}}$ was also detected for the first three years, but was later undetectable due to a low SNR (Figure \ref{fig:dynamic}). 
We did not include this component in the following analysis. 

\begin{figure*}[ht]
    \centering
    \includegraphics[width=0.9\linewidth]{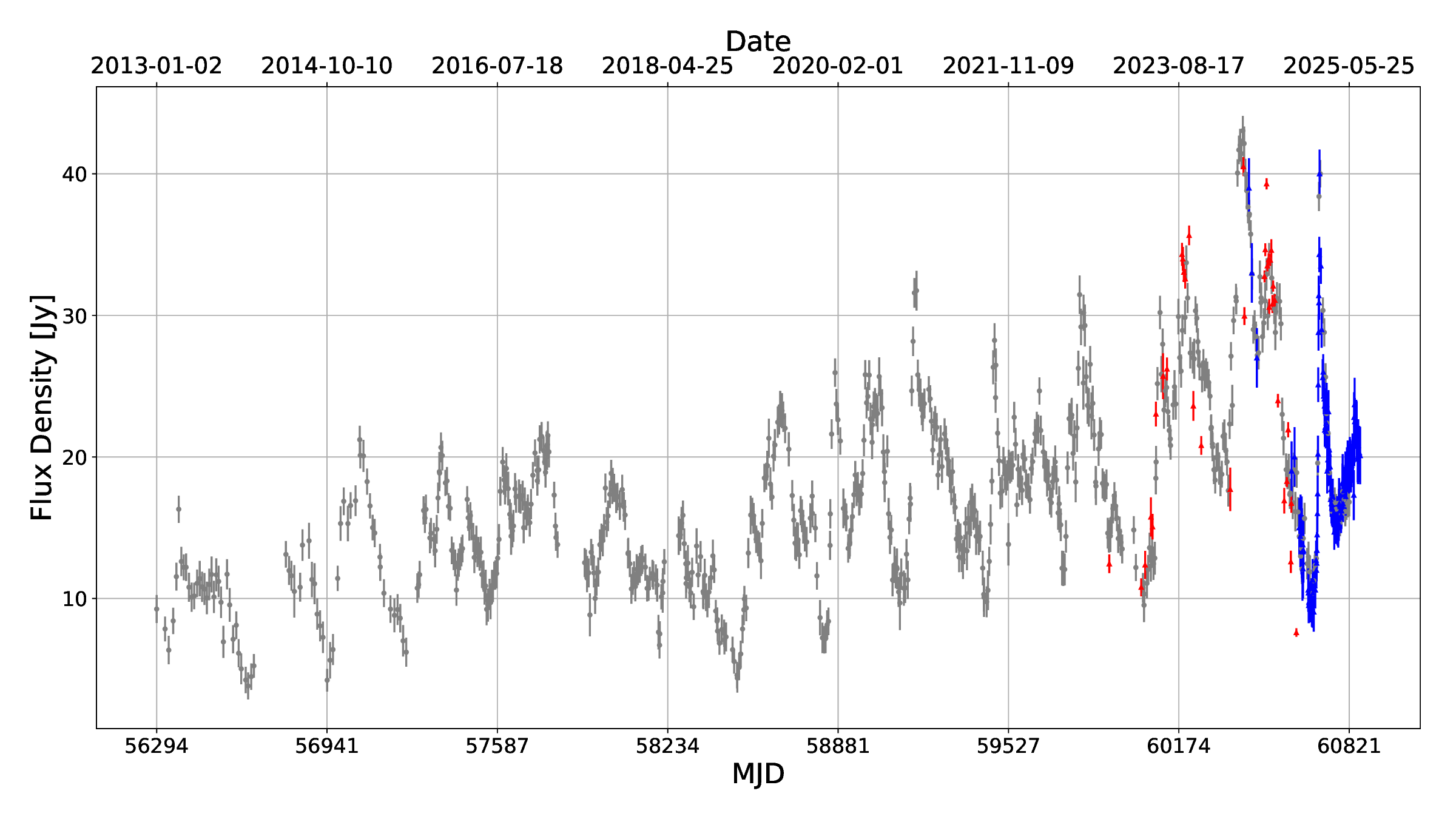}
    \caption{Light curve of the 6.7 GHz $\mathrm{CH_3OH}$ maser at $V_\mathrm{LSR}=-4.5\ \mathrm{km\ s^{-1}}$ for G353. 
    Gray, blue, and red points represent results from the Hitachi, Yamaguchi, and Tianma telescopes \citep{Song_2025ApJ...980..132S}, respectively. 
    Error bars indicate the 3-$\sigma$ noise level for each observation. 
    }
    \label{fig:lightcurve}
\end{figure*}

Figure \ref{fig:lightcurve} shows the light curve of the main velocity component ($V_\mathrm{LSR}=-4.5\ \mathrm{km\ s^{-1}}$).
We adopted the peak fluxes derived from Gaussian fitting for the Hitachi and Yamaguchi data. 
\citet{Song_2025ApJ...980..132S} conducted 36 epochs over $\sim$710 days during in 2022--2023.
They reported a possible periodicity of $\sim$330 days in the 6.7 GHz $\mathrm{CH_3OH}$ maser flux. 
We also identified a periodic variations of the maser, but the period was shorter than 330 d. 

\subsection{Method of the period analysis}
To quantitatively evaluate the quasi-periodic light curve in G353, we performed period analysis using the Lomb-Scargle (LS) periodogram method \citep{Lomb_1976Ap&SS..39..447L,Scargle_1982ApJ...263..835S} and the asymmetric power function (APF) fitting \citep{Szymczak_2011A&A...531L...3S}.

The LS method is suitable for analyzing irregular time-series data.
In this study, a period was considered significant when the p-value was less than 0.0001.
This hypothesis that the detected period arose by chance, corresponds to a confidence level of 99.99\% \citep{Frescura_2008MNRAS.388.1693F}.
The error of periods obtained by the LS method is estimated as the half width at half maximum of each peak in the LS Power.

The APF used in this study is defined as follows \citep{Szymczak_2011A&A...531L...3S,Tanabe_2024PASJ...76..426T}:
\begin{equation}
    S\left(t\right)=
    A \exp{s\left(t\right)}+Ct+D
\end{equation}
\begin{equation}
    s\left(t\right)=
    \frac{
    -B\cos{\left(2\pi \frac{t}{P}+\phi\right)}
    }{
    1-f\sin{\left(2\pi\frac{t}{P}+\phi\right)}
    }.
\end{equation}
Here, $A$, $C$, and $D$ are constants; $B$ represents the amplitude to the mean value; $P$ is the period; $\phi$ is the phase at time $t=0$; and $f$ represents the asymmetry parameter.
The parameter $f$ is the increase time from the minimum to the maximum divided by the period, and it ranges from $-1<f<1$ \citep{Szymczak_2011A&A...531L...3S}.
In this paper, we consider the function to be symmetric when $-0.5\leq f\leq 0.5$, following the definition by \citet{Tanabe_2024PASJ...76..426T}.

\begin{figure}
    \centering
    \includegraphics[width=1.1\linewidth]{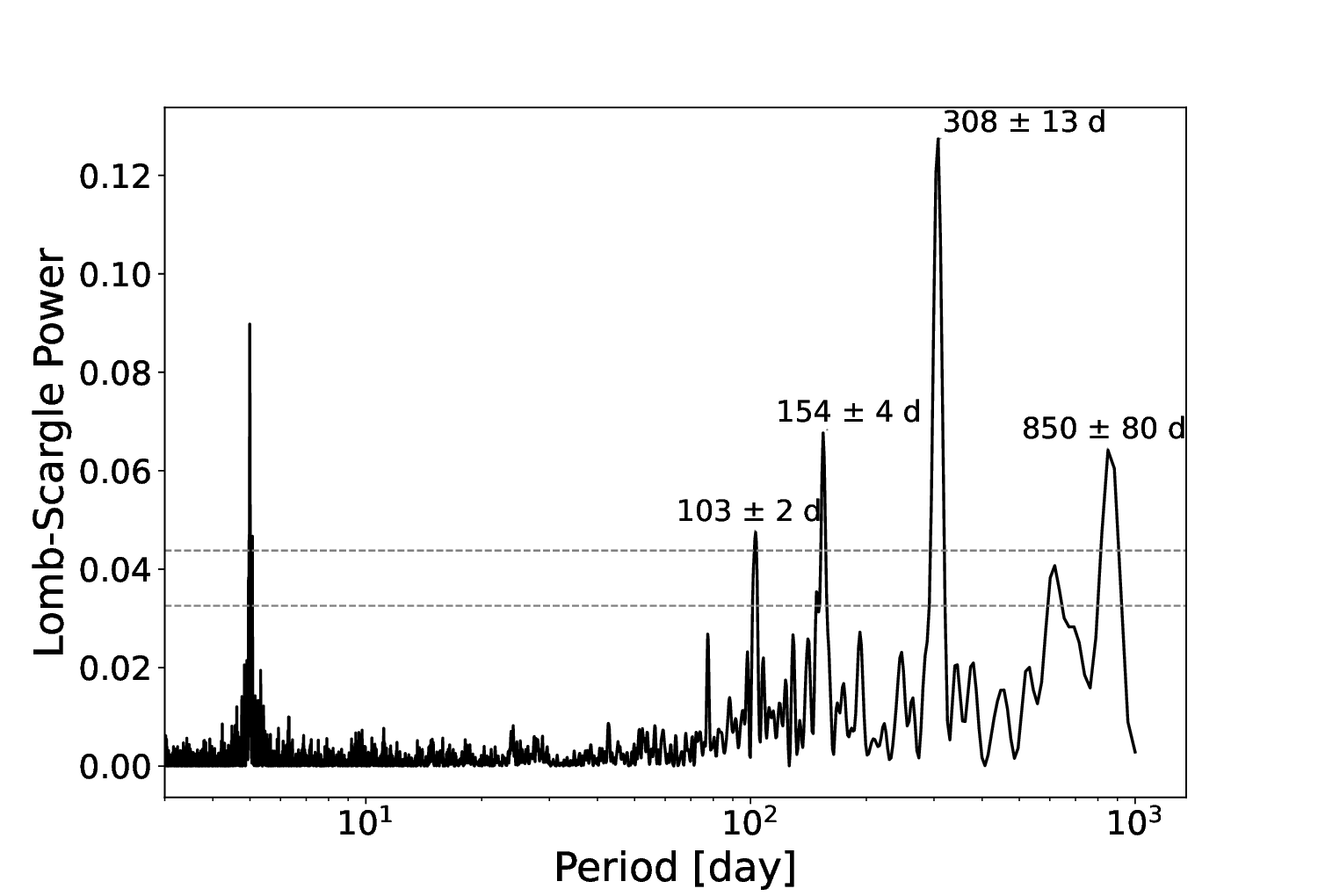}
    \caption{The results of the LS analysis at $V_\mathrm{LSR}=-4.5\ \mathrm{km\ s^{-1}}$.
    The dashed lines represent the p-values of 0.01 and 0.0001 from the bottom.}
    \label{fig:LS}
\end{figure}

Figure \ref{fig:LS} shows the LS Power of the peak flux at $V_\mathrm{LSR}=-4.5\ \mathrm{km\ s^{-1}}$. 
The sharp peak at 5 days was caused by the observation sampling cadence and was excluded because it did not represent a true periodicity. 
The LS analysis revealed a strong peak at a period of $308 \pm 13$ d.
The APF analysis yielded a period of $309.29\pm 0.23$ d, with an asymmetric profile.
Although multiple periods (including harmonics of the main peak) were detected, we focused only on the periods identified by both the LS analysis and the APF fitting. 

\begin{figure*}
    \centering
    \includegraphics[width=0.9\linewidth]{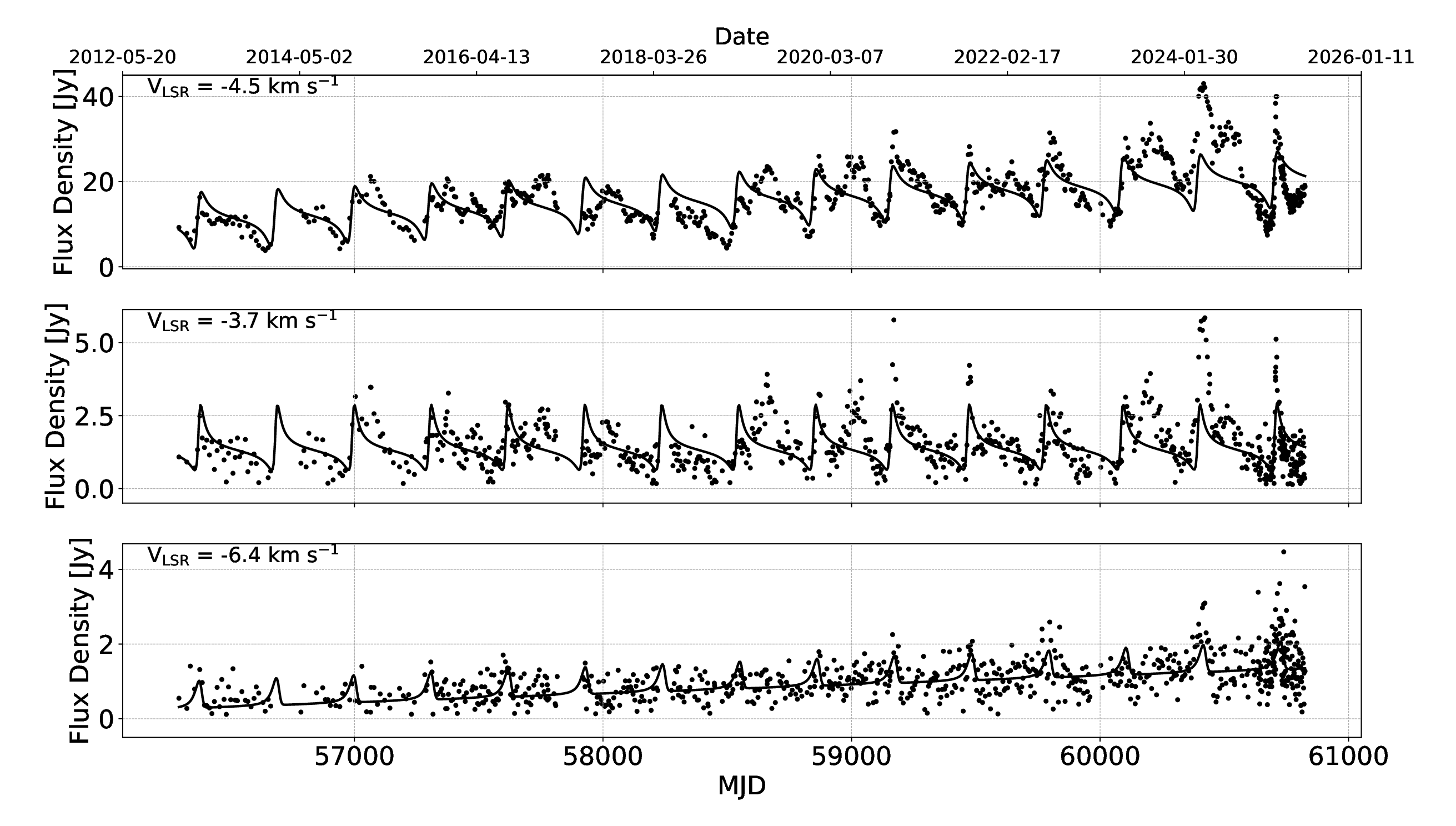}
    \caption{Light curves of the 6.7 GHz $\mathrm{CH_3OH}$ maser with the APF overlaid. 
    From top to bottom, the light curve for the velocity component of $V_\mathrm{LSR}=-4.5, -3.7, $ and $-6.4\ \mathrm{km\ s^{-1}}$ are shown, respectively. 
    }
    \label{fig:power_function}
\end{figure*}

The same analyses were also performed for the velocity components $V_\mathrm{LSR}=-3.7$ and $-6.4\ \mathrm{km\ s^{-1}}$ identified in Figure \ref{fig:average_spec}, with results summarized in Table \ref{tab:periods}.
Figure \ref{fig:power_function} shows the light curve for each velocity component, overlaid with the fitted APF.
In this analysis, the fitting was performed on the entire light curve without amplitude scaling in each cycle, i.e., the model sets the single best-fit amplitude parameter $A$. 
Hence, the model does not perfectly match the peak amplitude in each cycle but successfully reproduces the typical shape of the variability profile.
The APF profile for the $V_\mathrm{LSR}=-6.4\ \mathrm{km\ s^{-1}}$ component is different from the other two, probably due to the limited SNR. 

The periods of the $-4.5$ and $-3.7\ \mathrm{km\ s^{-1}}$ components agree with each other within the errors and their variations are synchronized.
In contrast, the period of the $-6.4\ \mathrm{km\ s^{-1}}$ component differs by $\sim1.5$ days from that of the other two components and this determines a time delay that becomes equal to $\sim2$ weeks at the end of our monitoring.

The light curve shows a characteristic asymmetric profile with a steep rise and a shallow decay. We regard the rising phase as the beginning of each cycle, i.e., $t=0$ corresponds to $\phi=0$ in Equation (2).
Irregular brightenings are seen in some cycles that do not align with the established periodicity, as shown in Figure \ref{fig:power_function}.

In addition to the periodic pattern, there is a long-term increasing trend for the $V_\mathrm{LSR}=-4.5$ and $-6.4\ \mathrm{km\ s^{-1}}$ components. 
Such a long-term trend is often seen for periodic maser sources \citep[e.g.,][]{Tanabe_2024PASJ...76..426T}. This may reflect an increasing trend of seed photons for maser amplification, since no such trend was seen in MIR pumping radiation (see next subsection).

\begin{deluxetable}{cccc}
\tablecaption{Results of Period Analysis \label{tab:periods}}
\tablehead{
\colhead{$V_{\mathrm{LSR}}$} & 
\colhead{$P_{\mathrm{LS}}$} & 
\colhead{$P_{\mathrm{APF}}$} & 
\colhead{$|f|$} \\
\colhead{(km\,s$^{-1}$)} & 
\colhead{(days)} & 
\colhead{(days)} & 
\colhead{}
}
\startdata
$-4.5$ & 308$\pm$13  & 309.29$\pm$0.23 & 0.96 \\
$-3.7$ & 308$\pm$15  & 309.28$\pm$0.18 & 0.97 \\
$-6.4$ & 312$\pm$13  & 310.97$\pm$0.34 & 0.86 \\
\enddata
\end{deluxetable}

\begin{figure}
    \centering
    \includegraphics[width=1\linewidth]{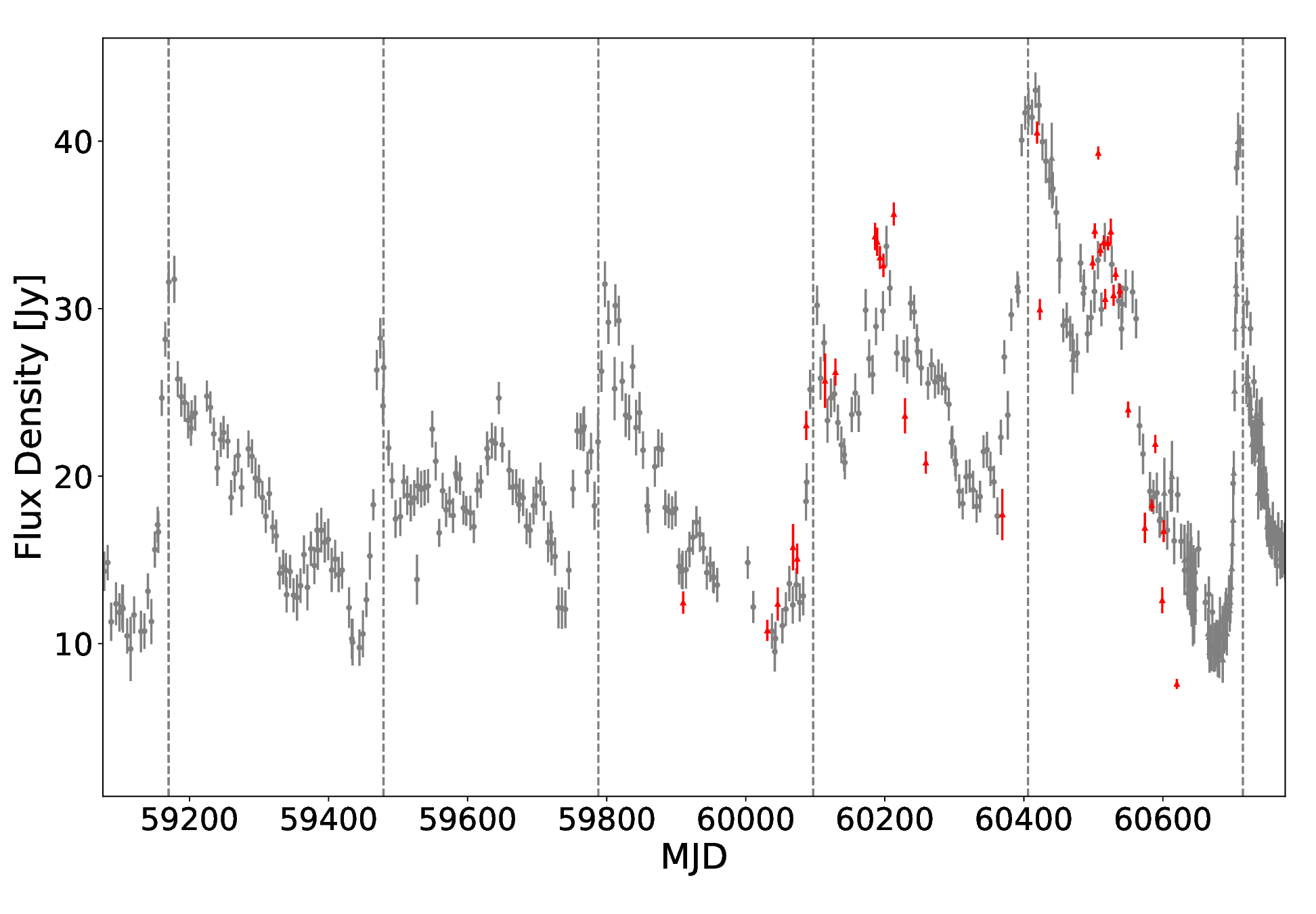}
    \caption{Enlarged view of the light curve shown in Figure \ref{fig:lightcurve}, focusing on the duration discussed by \citet{Song_2025ApJ...980..132S}. The gray dots represent data from the Hitachi and Yamaguchi radio telescopes, while the red dots represent data from the Tianma telescope. The dashed vertical lines indicate intervals of 309 d, corresponding to the maser cycle}.
    
    \label{fig:TMRT_lightcurve}
\end{figure}

Figure \ref{fig:TMRT_lightcurve} shows an enlarged view of the light curve in Figure \ref{fig:lightcurve}, comparing with the results of \citet{Song_2025ApJ...980..132S}. 
They reported periods of $320\pm 46$ days and $331\pm 4$ days based on the LS method and APF fitting, respectively.
Because of our longer monitoring duration and higher cadence, we could identify the accurate period and APF profile we detected.  
They treated the primary peak as an outlier in their profile fitting and mistakenly identified a secondary peak as the main one. This secondary peak sometimes appears and is inconsistent with the actual period derived from the LS analysis. We will discuss this sub-peak in detail in Section \ref{sec:discussion}. 

As a result, both the derived period and the variation profile in \citet{Song_2025ApJ...980..132S} differed from those obtained in our study. 
In particular, the asymmetry parameter reported by \citet{Song_2025ApJ...980..132S} was 0.35, while 0.96 in our case, differing by a factor of approximately three. 
Therefore, we consider the period of 309 days at $V_\mathrm{LSR}=-4.5\ \mathrm{km\ s^{-1}}$, obtained from the multiple cycles, as the most reliable. We proceed with our discussions adopting this period. 
We note that these discrepancies in period and asymmetry parameter do not significantly affect the discussion of thermal $\mathrm{CH_3OH}$ emission in \citet{Song_2025ApJ...980..132S}. 

\subsection{Infrared archival data by NEOWISE} 
\begin{figure*}
    \centering
    \includegraphics[width=0.9\linewidth]{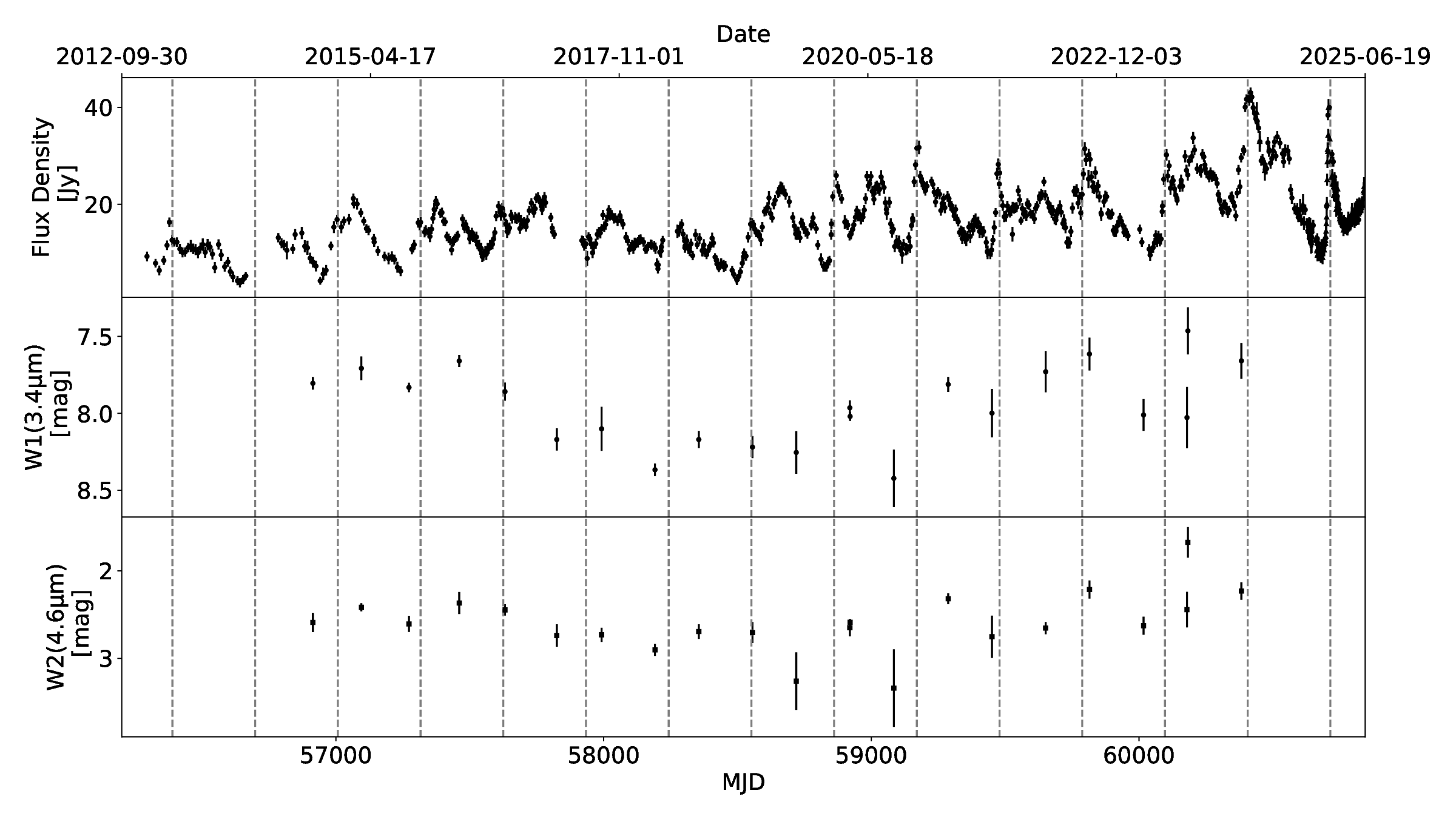}
    \caption{
    Comparison of the Light curves between the 6.7 GHz $\mathrm{CH_3OH}$ maser at -4.5 km s$^{-1}$ and two NEOWISE bands. 
    Error bars show 3-$\sigma$ noise level for the maser and standard error for the infrared data. The dashed vertical lines indicate intervals of 309 d.}
    \label{fig:neowise_lightcurve}
\end{figure*}
\begin{figure*}
    \centering
    \includegraphics[width=0.8\linewidth]{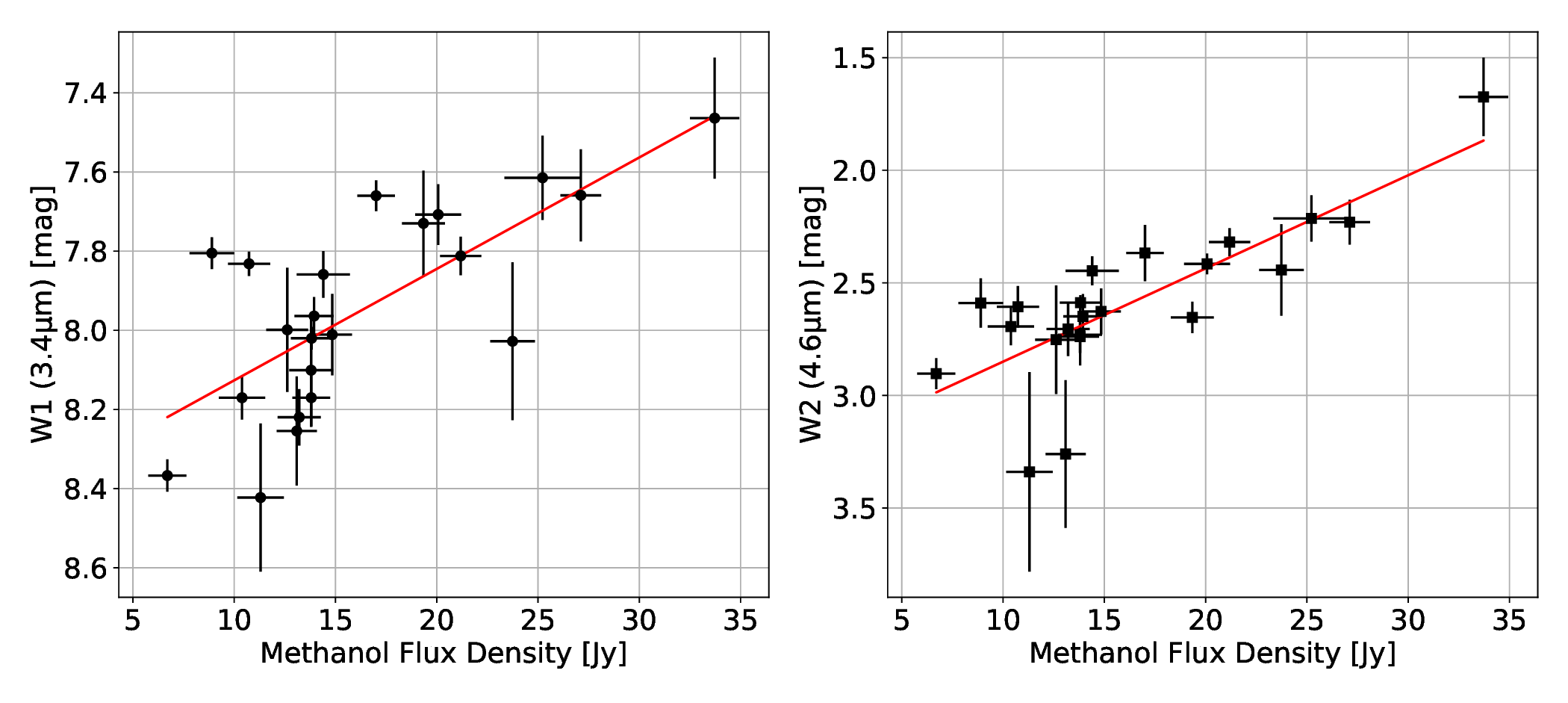}
    \caption{Correlation plots between the 6.7 GHz $\mathrm{CH_3OH}$ maser flux density and the NEOWISE W1 and W2 band magnitudes. 
    Error bars are given in the same definition as Figure \ref{fig:neowise_lightcurve}. 
    The red line represents a linear fit by the least-squares method. 
    }
    \label{fig:correlation}
\end{figure*}
Figure \ref{fig:neowise_lightcurve} compares the light curves of the 6.7 GHz $\mathrm{CH_3OH}$ maser and NEOWISE archival data.
When comparing the variations of the maser and MIR within each cycle, the $\mathrm{CH_3OH}$ maser and MIR light curves show consistent trends, that is, a decreasing trend observed in almost all cycle.
Figure \ref{fig:correlation} shows the correlation plots between the flux density of the 6.7 GHz $\mathrm{CH_3OH}$ maser and the magnitude of NEOWISE W1 and W2 bands. 
Each point corresponds to the maser and infrared data observed within $\pm 5$ days of each other.
The correlation coefficients for W1 and W2 are 0.72 and 0.78, respectively, indicating a positive correlation between the maser and these MIR magnitudes. 
\section{Discussion}\label{sec:discussion}

\subsection{Origin of 6.7 GHz $CH_3OH$ maser periodicity}
The theoretical models for periodic variability of 6.7 GHz $\mathrm{CH_3OH}$ masers can be categorized into five groups:
(1) Protostellar pulsation instability \citep{Inayoshi_2013ApJ...769L..20I},
(2) CWB \citep{vanderWalt_2011AJ....141..152V},
(3) Eclipsing binary \citep{Maswanganye_2015MNRAS.446.2730M},
(4) Spiral shock \citep{Parfenov_Sobolev_2014MNRAS.444..620P},
(5) Periodic accretion \citep{Artymowicz_Lubow_1996ApJ...467L..77A, Munoz_Lai_2016ApJ...827...43M}.
The models (2) to (4) are excluded in G353 for following reasons. 
First, G353 does not possess an H{\sc ii} region \citep{Motogi_2017ApJ...849...23M}, making the CWB model inappropriate, which presupposes the presence of an H{\sc ii} region. 
Second, G353 possesses a face-on disk.
Consequently, the eclipsing binary model and the spiral shock model, both of which require an edge-on disk configuration, are not applicable.

The model (5) is also unlikely, i.e., the periodic accretion model is characterized by recurring bursts and quiescent phases, as shown by observations from \citet{Araya_2010ApJ...717L.133A}. 
This behaviour is different from that of G353, which has a continuous light-curve profile with no quiescent phase. 
Moreover, in the interpretation proposed by \citet{Araya_2010ApJ...717L.133A} the accretion rate should vary from cycle to cycle, in contrast with the reproducibility of the asymmetric shape of the light curve observed in G353.

Moreover, models (2) to (5) are premised on being caused by binary stars or circumbinary disks.
However, as of now, G353 shows no signs of a compact binary system, even in the long-baseline observations by ALMA(spatial resolution $\sim$ 20--30 milli-arcsecond, corresponding to $\sim$ 34--51 au).
In contrast, the protostellar pulsation model proposed by 
\citet{Inayoshi_2013ApJ...769L..20I} can consistently explain the periodic variability observed in G353 (see later discussions). 

The superradiance scenario proposed by \citet{Rashidi_2025MNRAS.542L..12R} also explains burst-like brightening. It could produce a dip-like profile in some cases \citep{Rajabi_2023MNRAS.526..443R}. However, this model predicts velocity-dependent differences in flare duration, but no clear differences are observed in our data. In addition, the dip-like feature appears after the flare in this model, contrary to the case of G353. 
Therefore, the present data do not match the superradiance model.

\subsection{Correlated variability between MIR and maser}

We confirmed a time-series correlation between the MIR and $\mathrm{CH_3OH}$ maser emission in G353, as shown in Figure \ref{fig:correlation}. 
\citet{Uchiyama_2022ApJ...936...31U}  reported a similar correlation between the NEOWISE data and the 6.7 GHz $\mathrm{CH_3OH}$ maser flux in a high-mass protostar G036.70+00.09. 
It is the first confirmation of a MIR-maser correlation in a periodic 6.7 GHz $\mathrm{CH_3OH}$ maser source, while such a correlation has previously been reported as a transient phenomenon associated with the accretion burst in S255-NIRS3 \citep{CarattioGaratti_2017NatPh..13..276C}.
\citet{Durjasz_2019MNRAS.485..777D} also reported a strong MIR-maser correlation for a highly variable but non-periodic 6.7 GHz $\mathrm{CH_3OH}$ maser source. 

Such correlations are expected from the well-known IR pumping model in \citet{Cragg_2005MNRAS.360..533C}, and a statistical trend that Class II $\mathrm{CH_3OH}$ maser luminosity correlates with bolometric luminosity \citep[e.g.,][]{Bartkiewicz_2014A&A...564A.110B,Jones_2020MNRAS.493.2015J}. 
In addition to the standard pumping model assuming the MIR re-emission by heated dusts, \citet{Zinchenko_2025MNRAS.543L...9Z} have recently reported that direct MIR radiation from the host protostar can excite some class II $\mathrm{CH_3OH}$ masers at a millimeter band. 
Such a direct excitation can also be applicable for 6.7 GHz $\mathrm{CH_3OH}$ maser in principle. 

The light curve of the $\mathrm{CH_3OH}$ maser in G353 reflects the variability of the protostellar luminosity directly, rather than flux variations of the maser seed photons. 
The protostellar pulsation model naturally explains such a periodic variation in the (Mid-)IR luminosity. 

\subsection{Maser variation profiles for each cycle}
A detailed profile of a light curve provides an important clue to discuss the physical origin of the periodicity. 
This fact suggests the involvement of multiple factors in the $\mathrm{CH_3OH}$ maser variability of G353.

\begin{figure*}
    \centering
    \includegraphics[width=0.8\linewidth]{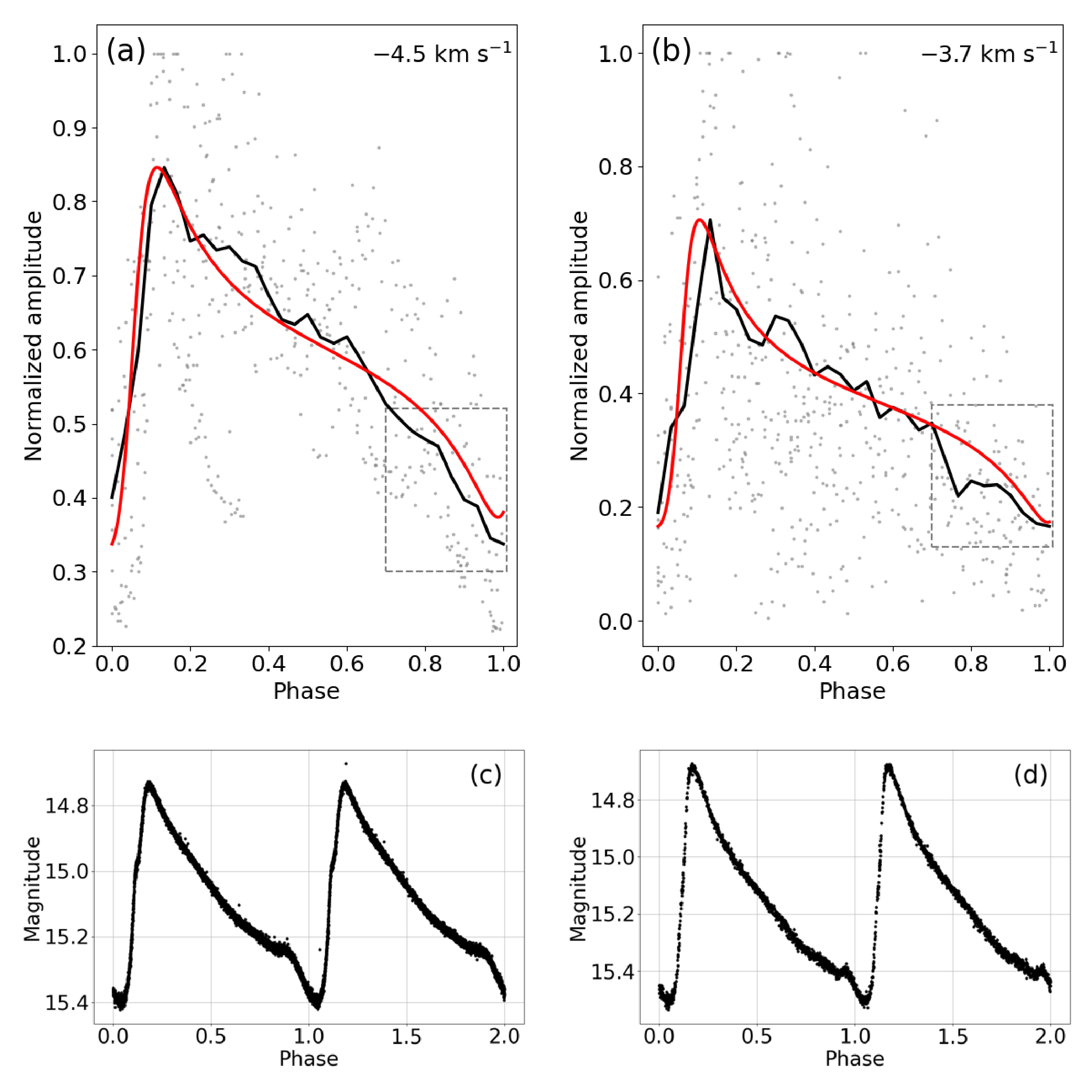}
     \caption{(a) and (b): Black line displays the averaged profiles of the light curve at $V_\mathrm{LSR}=-4.5$ and $V_\mathrm{LSR}=-3.7$ $\mathrm{km\ s^{-1}}$.
     We calculated the average for every phase bins of 0.03 in each period. 
     The red line shows the overlaid Asymmetric Power Function (APF) obtained in Section 4.2. Gray points represent the original data points before averaging. The dashed rectangle roughly indicates the phase range displaying the steep drop in intensity. (c) and (d) show examples of a light curve for RR Lyrae and Anomalous Cepheids at I-band, respectively \citep{Soszynski_2017AcA....67..297S}. These are folded and overlaid for every two periods.}
    \label{fig:Average_profile}
\end{figure*}

In Figure \ref{fig:Average_profile}(a) and (b), black line displays the typical profile constructed by averaging a normalized profile for each period at $V_\mathrm{LSR}=-4.5$ and $V_\mathrm{LSR}=-3.7$ $\mathrm{km\ s^{-1}}$, excluding five cycles that show clear secondary peaks in their light curves. 
The red line, which represents the APF obtained in Section 4.2, is overlaid, confirming the shape of the averaged profile.
Figure \ref{fig:Average_profile}(c) and (d) show examples of a light curve of pulsating variable stars, specifically RR Lyrae and Anomalous Cepheids \citep{Soszynski_2017AcA....67..297S}.
The averaged profile of G353 exhibits good agreement with these characteristic light curves of pulsating variable stars.
Furthermore, the amplitudes of maser variation are comparable to those found in such stars.

A particularly noteworthy feature of the averaged light curve is the steep drop in intensity, which is observed immediately before brightening. 
We highlight this feature by the dashed rectangle in Figure \ref{fig:Average_profile}(a) and (b).
Hereafter we simply call this feature as dip.
Such a dip feature is seen in the light curve of some pulsating variables as in Figure \ref{fig:Average_profile}(c) and (d). 
This feature is thought to arise from a stellar internal ionization region that controls the energy transport process via convection \citep{Bono_Stellingwerf_1994ApJS...93..233B}.
Although it remains unclear whether the internal structure of the protostar truly matches that of pulsating variable stars, dip in the light curve is at least distinct from a simple exponential cooling following a non-steady heating process such as periodic accretion \citep{Francis_2022ApJ...937...29F}. 
Therefore, we suggest that the observed dip structure is an indicative sign supporting the hypothesis that G353 is undergoing protostellar pulsation. 

\subsection{Physical parameters assuming pulsation instability}
If the periodic pulsation of the protostar itself is the primary cause of the $\mathrm{CH_3OH}$ maser variability observed in G353, we can deduce several protostellar parameters based on theoretical models.
\citet{Inayoshi_2013ApJ...769L..20I} proposed that protostars become pulsationally unstable during the pre-main-sequence expansion phase.
In this pulsationally unstable state, the periodic variation in protostellar luminosity causes the temperature of the surrounding dust also to vary periodically, leading to a periodic change in the maser flux. 
The pulsation timescale of a protostar can be estimated by the free-fall time, $t_\mathrm{ff}\sim \sqrt{1/G\rho}$.
Substituting the observed period of 309 d, the expected density is $\sim2\times 10^{-5}\ \mathrm{kg\ m^{-3}}$($\sim6\times 10^{15}\ \mathrm{cm^{-3}}$ in number density).
The protostellar mass of G353 is estimated to be $\sim10\ M_\odot$ from its luminosity \citep{Motogi_2017ApJ...849...23M}.
A simple calculation yields a protostellar radius of $880\ R_\odot$ from an averaged density. 
This suggests a highly extended, red supergiant(RSG)-like atmospheric structure.
Such extended stars typically have a density gradient, with density decreasing exponentially with radius. 
When applying the typical density gradient of a RSG, it has been reported that the density of most of the stellar atmosphere is about 20\% of an averaged density \citep{Kozyreva_2020MNRAS.494.3927K}.
Considering this, the radius of G353 would be $\sim500\ R_\odot$.

Meanwhile, \citet{Inayoshi_2013ApJ...769L..20I} proposed the following equations to estimate the protostar's radius, mass accretion rate, and luminosity from the variability period:
\begin{equation}
    R_* = 350\ R_\odot
    \left(
    \frac{P}{100\ \mathrm{d}}
    \right)^{0.62}
\end{equation}
\begin{equation}
    \dot{M_*} = 3.1\times 10^{-3}\ M_\odot \mathrm{yr}^{-1}
    \left(
    \frac{P}{100\ \mathrm{d}}
    \right)^{0.73}
\end{equation}
\begin{equation}
    \log_{10}{\left( \frac{L_*}{L_\odot}\right)} = 4.62 + 0.98\log_{10}{\left( \frac{P}{100\ \mathrm{d}}\right)}
\end{equation}

\begin{deluxetable*}{lccc}
\tablecaption{Protostellar parameters of G353 derived from $t_\mathrm{ff}$ and pulsation models \label{tab:parameters_G353}}
\tablehead{
    \colhead{Method} & \colhead{Mass Accretion Rate} & \colhead{Radius} & \colhead{Luminosity}
}
\startdata
Estimation with $t_\mathrm{ff}$     &  & $500~R_\odot$ &  \\
Pulsation Model    & $6 \times 10^{-3}~M_\odot~\mathrm{yr}^{-1}$ & $700~R_\odot$ & $10^5~L_\odot$ \\
Observed Values    & $3 \times 10^{-3}~M_\odot~\mathrm{yr}^{-1}$ &  & $>5500,\ 14000~L_\odot$ \\
\enddata
\end{deluxetable*}

Table \ref{tab:parameters_G353} summarizes the results calculated by substituting the period in these equations.
For the observed values, the mass accretion rate is adopted from \citet{Motogi_2019ApJ...877L..25M}, while the protostellar luminosity values are taken from 
\citet{Motogi_2017ApJ...849...23M}, and \citet{Urquhart_2022MNRAS.510.3389U}.
Regarding the mass accretion rate, the observed value of $3\times 10^{-3}\ M_\odot\mathrm{yr}^{-1}$ obtained by \citet{Motogi_2019ApJ...877L..25M} is consistent with the pulsation model within a factor of 2.
The protostellar radius also agrees with the rough estimation based on the free-fall time.
The protostellar luminosity shows a significant difference, being approximately 20 times higher than the minimum estimate of $5500\ L_\odot$ \citep{Motogi_2017ApJ...849...23M}.
Although there is some deviation in the reported luminosities depending on the SED model, the luminosity is still one order of magnitude lower than the theoretical model, even for the most luminous estimate ($14000\ L_\odot$). 

\begin{figure*}
    \centering
    \includegraphics[width=0.8\linewidth]{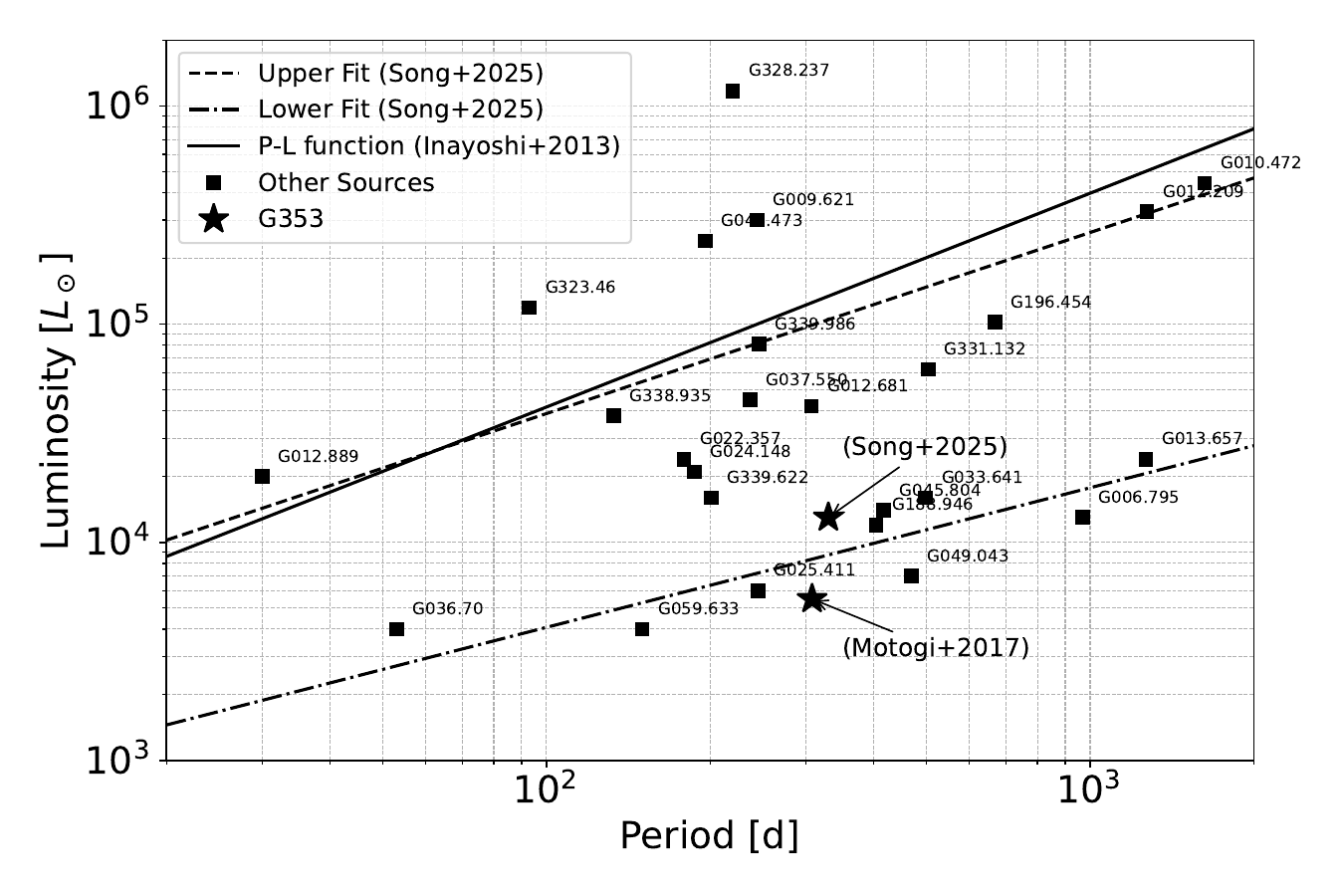}
    \caption{Period-luminosity relation of periodically variable 6.7 GHz $\mathrm{CH_3OH}$ maser sources analyzed by \citet{Song_2025ApJ...980..132S}. Black squares indicate the periodically variable maser sources, with the star representing G353. Luminosity values from \citet{Song_2025ApJ...980..132S} and \citet{Motogi_2017ApJ...849...23M} are plotted separately due to the difference of the luminosity. The black solid line represents the PL function from \citet{Inayoshi_2013ApJ...769L..20I}, while the dotted lines indicate linear fits for the upper and lower groups derived by \citet{Song_2025ApJ...980..132S}.}
    \label{fig:PLfunction}
\end{figure*}

This pulsation model is calculated based on the spherically symmetric accretion condition by \citet{Hosokawa_Omukai_2009ApJ...691..823H}.
It has been pointed out that even with the same accretion rate, the maximum expansion radius of a protostar can differ significantly depending on the initial entropy distribution
\citet{Stahler_a_1980ApJ...242..226S}, \citet{Stahler_b_1980ApJ...241..637S}, and \citet{Haemmerle_2016A&A...585A..65H}.
\citet{Song_2025ApJ...980..132S} proposed that high-mass protostars in earlier evolutionary stages tend to exhibit lower luminosity than expected from models in their period-luminosity relation based on a statistical study. 
Figure \ref{fig:PLfunction} presents the period-luminosity relation of periodically variable $\mathrm{CH_3OH}$ maser sources statistically analyzed by \citet{Song_2025ApJ...980..132S}.
G353 is located significantly below the luminosity predicted by the theoretical equation of \citet{Inayoshi_2013ApJ...769L..20I}, but it generally aligns with the linear fit for the lower group, which is considered to be in an early evolutionary stage.

We calculated a protostellar surface temperature for the three luminosity estimations (5500, 14000 $L_\odot$) by using the Stefan-Boltzmann law. 
The estimated effective temperatures are $T_\mathrm{eff}=2200$ and 2800 K, respectively. Here we subtracted the accretion luminosity ($L_\mathrm{acc}\sim GM\dot{M}/R$) from the total luminosities, assuming the observed accretion rate. 
These low temperatures could imply a cool protostellar atmosphere due to its remarkably expanded radius \citep{Meyer_2019MNRAS.484.2482M}. 

\subsection{Irregular brightening features}
We consider that protostellar pulsation can explain most of the features of the maser periodicity in G353. 
On the other hand, the irregular brightening events that do not align with the cycle could be caused by non-periodic phenomena independent of protostellar pulsation. 
For example, episodic accretion events and/or episodic jet driving events can happen. 
The unstable accretion disk reported by \citet{Motogi_2019ApJ...877L..25M} implies that the episodic accretion potentially occurs in G353. 
Moreover, episodic jet activity has also been reported in G353 at intervals of 1--2 yr \citep{Motogi_2016PASJ...68...69M}. 
Such a jet-driving event can, for instance, be accompanied by brightening due to shock heating. 
Alternatively, variations in the infrared optical depth near the protostar due to uplifted gas/dust can also cause apparent luminosity variations. 
Simultaneous monitoring observations of the $\mathrm{CH_3OH}$ and $\mathrm{H_2O}$ maser will allow us to verify the relationship between the irregular peaks and such jet activity. 

\subsection{Another remaining possibility: protobinary scenario}
Although the pulsation model would successfully explain observed features, we cannot rule out the periodic accretion scenario caused by an unresolved protobinary as the origin of maser periodicity entirely.
If we assume a protobinary scenario, certain observational constraints must be satisfied.
First, the association of a Class II 6.7 GHz $\mathrm{CH_3OH}$ maser and the high bolometric luminosity requires that at least a part of the protobinary is a high-mass protostar. 
Second, the total mass of the protobinary system must be consistent with the observed bolometric luminosity.
Third, given that G353 has a face-on geometry, orbital eccentricity is required to produce the observed periodicity. 
Such an eccentric binary interacts only at the periastron, producing an adequate time interval. 
The allowable total mass of the binary system is not uniquely constrained, since the estimated luminosities of G353 varied by a factor of two depending on the adopted SED models. 
In this study, we consider two representative model cases: the protobinary system consisting of 10 $M_\odot$ and 1 $M_\odot$ objects, and the equal-mass protobinary with two 8 $M_\odot$ objects.

\begin{deluxetable*}{cccccccc}
\tablecaption{Binary Parameters and Angular Separations for Two Modeled Systems \label{tab:binary_models}}
\tablehead{
\colhead{} & \colhead{} & \multicolumn{3}{c}{10~$M_\odot$ : 1~$M_\odot$} & \multicolumn{3}{c}{8~$M_\odot$ : 8~$M_\odot$} \\
\colhead{Orbital Phase} & \colhead{} & \colhead{$e=0.3$} & \colhead{$e=0.5$} & \colhead{$e=0.7$} & \colhead{$e=0.3$} & \colhead{$e=0.5$} & \colhead{$e=0.7$}
}
\startdata
Periastron & Separation [au]           & 1.39 & 0.99 & 0.60 & 2.59 & 1.13 & 0.68 \\
           & Angular separation [mas]  & 0.82 & 0.58 & 0.35 & 1.52 & 0.66 & 0.40 \\
Apastron   & Separation [au]           & 1.58 & 2.98 & 3.38 & 2.93 & 3.38 & 3.83 \\
           & Angular separation [mas]  & 0.93 & 1.75 & 1.99 & 1.72 & 1.99 & 2.25 \\
\enddata
\end{deluxetable*}
For each case, we assumed a plausible eccentricity and calculated the angular separations (Table \ref{tab:binary_models}). 
These calculations assumed a distance of 1.7 kpc. 
The calculated binary separations at apastron correspond to angular distances of 0.93 to 2.25 milli-arcsecond (mas). 
These values are below the highest resolution of ALMA \citep[10--20 mas at Band 6, $\sim$5 mas at Band 10;][]{ALMA_partnership_2015ApJ...808L...3A}, making it impossible to resolve the binary separation. 
In the case of the equal-mass binary (or an adequately bright companion), we may identify phase-dependent shifts in the brightness profile within a synthesized beam along the orbital phase. It may provide an indirect signature of the protobinary if we measure such a sub-beam structure. 

The next-generation Very Large Array (ngVLA) is expected to achieve angular resolutions of about 1 mas and brightness temperature sensitivities of several thousand Kelvin in the 27--93 GHz band at 7$\sigma$ or better SNR \citep{Selina_2018ASPC..517...15S}. 
This extreme sensitivity and resolution will enable us to resolve a protobinary at apastron directly. 
On the other hand, if G353 is really a pulsating RSG-like protostar with a radius of $\sim$500 $R_\odot$ (diameter $\sim$4.6 au), its angular size would be about 2.7 mas, allowing for the potential of direct imaging of the protostellar radio photosphere \citep{ngVLAmemo_2021arXiv210308859T}. 
We are convinced that future high-resolution observations with ngVLA will provide a direct and conclusive confirmation of whether G353 is a single bloated protostar or a protobinary system. 

\section{Conclusions}
We conducted a long-term monitoring observation of the 6.7 GHz $\mathrm{CH_3OH}$ maser associated with the high-mass protostar G353 using the Hitachi and Yamaguchi 32m radio telescopes. 
We detected significant flux variations with a periodicity of $\sim$309 d. 
While several models have been proposed to explain periodic variability of 6.7 GHz $\mathrm{CH_3OH}$ masers, the absence of H{\sc ii} region and the presence of the face-on disk in G353 exclude most models involving binary interaction or circumbinary disks. 

Archival NEOWISE data revealed a positive correlation between the variations of the maser and MIR emission at 3.4 and 4.6 $\mu$m. 
This fact indicates that variations in MIR pumping radiation cause maser flux variations, likely reflecting the variable protostellar luminosity. 
The average light curve profile shows the dip just before the luminosity rise, which is a characteristic feature seen in some of pulsating variable stars. 
With these facts, we propose that the protostellar pulsation model is a plausible explanation for the observed maser periodicity in G353. 

Based on this, the protostellar density estimated from the free-fall timescale is $\sim2\times 10^{-5}$ $\mathrm{kg\ m^{-3}}$, with a protostellar radius of about 500 $R_\odot$.
This radius is broadly consistent with protostellar parameters predicted by the theoretical model proposed by \citet{Inayoshi_2013ApJ...769L..20I}.
Only the observed luminosity significantly differs from the theoretical prediction, which may be explained if G353 has a larger radius and a cooler atmosphere than the model assumed.
Additionally, there were several irregular sub-peaks in the light curve, which could be caused by non-steady accretion events or jet activity independent of the protostellar pulsation. 

Another explanation for the maser periodicity via the periodic accretion caused by an unresolved compact protobinary. Although no signature of such a protobinary has yet been identified with ALMA, we cannot completely rule out this scenario at present.
Extremely high-resolution imaging with next-generation interferometers such as the ngVLA will allow us to directly resolve a possible protobinary or radio photosphere of a bloated protostar. It will provide direct and conclusive confirmation of the protostellar pulsation scenario.

\begin{acknowledgments}
We appreciate the anonymous referee’s fruitful comments and suggestions.
This work is partially supported by the Inter-university collaborative project ``Japanese VLBI Network (JVN)" of NAOJ and by the Japan Society for the Promotion of Science (JSPS) KAKENHI Grant Numbers JP21H01120 and JP21H00032.
This publication makes use of data products from the Near-Earth Object Wide-field Infrared Survey Explorer (NEOWISE), which is a joint project of the Jet Propulsion Laboratory/California Institute of Technology and the University of California, Los Angeles. NEOWISE is funded by the National Aeronautics and Space Administration.
This research has made
use of the NASA/IPAC Infrared Science Archive (IRSA), which
is operated by the California Institute of Technology, under
contract with the National Aeronautics and Space Administration.
The archival IRSA services and data include the NEOWISE-R Single
Exposure table \citep{NEOWISE_https://doi.org/10.26131/irsa144}.
\software{Matplotlib \citep{Matplotlib_2007CSE.....9...90H}, Numpy \citep{Numpy_2020Natur.585..357H}, Astropy \citep{Astropy_2022ApJ...935..167A}}

\end{acknowledgments}

\bibliography{Harajiri_2025}{}
\bibliographystyle{aasjournalv7}



\end{document}